\documentclass[12pt,preprint]{aastex}

\usepackage{natbib}

\begin{document}


\title{The Masses Of The B-Stars In The High Galactic Latitude Eclipsing Binary IT Lib\altaffilmark{1}}


\author{John C. Martin}
\affil{Warner \& Swasey Observatory, Case Western Reserve University,\\
    Cleveland, OH 44106-7215}
\email{martin@fafnir.cwru.edu}

\altaffiltext{1}{Based on observations made at
 the 2.1-m Otto Struve Telescope of McDonald Observatory operated by the University of Texas at Austin and
also at the 2.1m telescope at Kitt Peak National Observatory, National Optical Astronomy Observatories, operated by the Association of Universities for Research in Astronomy (AURA) under cooperative agreement with the National Science Foundation}


\begin{abstract}
A number of blue stars  
which appear to be similar to Population I B-stars in the star forming 
regions of the galactic disk are found more than 1 kpc from the galactic 
plane.  Uncertainties about the true distances and 
masses of these high latitude B-stars has fueled a debate as to their origin 
and evolutionary status.  The eclipsing binary IT Lib is composed of two 
B-stars, is approximately one kiloparsec above the galactic plane, and is 
moving back toward the plane.  Observations of the light and velocity curves 
presented here lead to the conclusion that the B-stars in this system are 
massive young main-sequence stars.  While there are several possible 
explanations, it appears most plausible that the IT Lib system formed in 
the disk about 30 million years ago and was ejected on a trajectory taking 
it to its present position.  
\end{abstract}




\section{Background}
Faint blue stars at high galactic latitude were first detected in a survey 
conducted by \citet{humason47}.  Soon after, \citet{bidelman51} brought 
attention to a number of high velocity stars of early spectral type.  The 
faint apparent magnitudes (sizable distances) and high velocities led to 
the conclusion that these objects, which appear at low spectroscopic 
dispersions to be Population I stars, were a part of the galactic halo.  
A debate has raged since about the origin and characteristics of these high 
galactic latitude B-stars.  Uncertainty about the masses or true distances 
of these stars has left the door open to many different scenarios.  
The discovery that \objectname[HD 138503]{IT Lib} is an eclipsing 
binary \citep{ibvs99} raises to three
the number of known eclipsing binaries at least 1 kpc above the galactic plane
whose components are B-type stars.  The other two are  V Tuc \citep{wood58} 
and DO Peg \citep{hoff35}.  
IT Lib is the brightest of the three and has light curve and radial velocity
data gathered for it.  As a result, IT Lib presents us with an 
excellent opportunity to directly measure the masses of a pair of high 
galactic latitude B-stars.

Photometric variability was first noted in 
\objectname[HD 138503]{IT Lib (HD 138503)} by 
\citet{stroh67}.  \citet{hill70} noted variability on the order of 0.2 
magnitudes.  \citet{mss88} classified the spectra of 
\objectname[HD 138503]{IT Lib (HD 138503)} as B2/B3 
(IV)(n) with a note that the lines appeared "somewhat filled in."  Later 
observations by the Hipparcos satellite confirmed 
\objectname[HD 138503]{IT Lib's (HIP 76161)} photometric 
variability.  The $74^{th}$ naming list of the IBVS \citep{ibvs99} 
assigned the designation \objectname[HD 138503]{IT Lib} and 
classified it as a possible eclipsing binary.

Analysis of the Hipparcos photometry annex data yielded an 
ephemeris with the parameters:
$P=2.2674600$ days, $T_0=2448500.76777$ JD   \citep{hip}.  
Unfortunately, there was a beat pattern between the period of IT Lib 
and the duty cycle of the Hipparcos satellite so substantial gaps were left 
in the phase coverage from phase 0.30 to 0.58 and phase 0.63 to 
0.75.  The Hipparcos data show a strong primary eclipse but 
the secondary eclipse falls in the gap of observations.  

\section{New Spectral Data}

High resolution spectra were taken of IT Lib with the Sandiford Echelle 
Spectrograph \citep{sandiford} on the McDonald Observatory 82" Struve 
Reflector in March 1999, February 2001, and May 2002.  A summary of the 
observations is given in Table 1.  The spectra were reduced 
and extracted using IRAF and analyzed using the ASP software package
of \citet{asp}.

\ion{H}{1} lines, \ion{He}{1} lines, and the \ion{Mg}{2} 4481 
line all have profiles with two well defined minima over most of the 
observed period (see Figure 1 for an example).  The broadening 
due to rotation of the individual components is on the order of 100-150 km/s 
so that the lines from each component never completely separate from each 
other.   This blending probably accounts for the line peculiarities noted 
by \citet{mss88}.  As modeled in section 4 of this paper, the primary 
component is three to four times as bright as the secondary in the band-pass of
these spectroscopic observations so that its spectral 
lines dominate the composite spectra.  An assortment of \ion{N}{2}, 
\ion{O}{2}, \ion{Al}{3}, \ion{Si}{3}, and \ion{Fe}{3} lines could be 
identified at the radial velocity of the primary component but no 
lines other than \ion{H}{1}, \ion{He}{1}, or \ion{Mg}{2} 4481 
could be positively identified and attributed to the secondary.  

The minima of the He I line profiles were used to measure the radial 
velocity of each component.  At some phases, the center of the contribution 
from the secondary could not be measured due to blending so that only the 
radial velocity of the primary was measured at those phases.  Table 2 
summarizes the radial velocity measurements made from the observations and 
Figure 2 shows them plotted by phase using the \citet{hip} parameters.

For the analysis in section 4 the individual weights of each radial velocity
measure are set equal to each other.  The measurements of the radial velocity
of the secondary are somewhat more uncertain than those of the primary's 
velocity because
the lines of the secondary are significantly weaker and often partially 
blended with the lines of the primary.  The relative weighting of the two 
velocity curves is accounted for in the curve 
dependent weights assigned to each data set as discussed in section 4.

\section{New Photometric Data}

The photometric data for IT Lib from the Hipparcos Catalog Epoch Photometry 
Annex has substantial gaps in phase space.  An effort was made to obtain 
more photometric observations of IT Lib at the unobserved phases in order 
to verify that the system does have a secondary eclipse and to provide more 
points for light curve fitting in order to determine the properties of the 
system.  Two sets of additional data were obtained.  \citet{feld02} obtained 
observations with the 2048x2048 Tektronix CCD (T2KA) with a Johnson B filter
on the Kitt Peak 2.1-m.  Those observations were reduced using sky flats and 
the non-linearity correction of \citet{mochejska01}.  The bulk of the CCD 
observations were obtained using the 1024x1024 FLICam (SIA-003 AB SITe 
back-illuminated CCD) using Cousins R and I filters on the 
Warner \& Swasey Observatory, Nassau Station 36" reflector.  Table 3 gives 
a summary of the observations.

The data were reduced in IRAF and aperture photometry was done with the 
IRAF procedure QPHOT.  The apertures for QPHOT were set to approximately 
3 sigma the FWHM of the stellar profiles.  Table 4 identifies the reference 
stars.  All the reference 
stars are between three and five magnitudes fainter than the target.  There 
are no brighter reference stars in the fields of view 
($10.4\arcmin\times10.4\arcmin$ for the 2.1-m data and 
$5\arcmin\times5\arcmin$ for the 36'' data).  As a result, 
uncertainty in the background turned out to be the largest contributing 
factor to the photometry errors. 

The simple average of the scaled relative magnitudes with respect to 
each reference star is taken to be the measured relative magnitude of IT 
Lib for each individual image. The sigma is taken as the random error of 
each individual measurement.  The errors are higher for the measurements 
in Cousins R because the night sky is significantly brighter in that 
band-pass.  Larger errors also occurred on nights where moon light and/or 
clouds and humidity made significant contributions to the sky brightness.  
Only data with errors smaller than 50 milli-magnitudes was 
used in the analysis (Figure 3).  

There was no need for absolute calibration of the photometry gathered.
The \citet{wilson01} code used to perform the analysis is capable of
performing a simultaneous solution to multiple light curves in multiple
filters/band-passes.  The relative magnitudes in each filter were converted to
relative intensity normalized to phase 0.25 and inputted as four separated 
data sets (Johnson B, Cousins R, Cousins I, and Hipparcos V) along with
the appropriate effective wavelength data for each filter.  Individual
data points were weighted by the inverse square of their error.

\section{Analysis of Orbits and Masses of Components}

The parameters of the IT Lib system were derived from the radial velocity 
and photometric data using the program of \citet{wilson01}.  
\citet{wilson01} uses the model of \citet{wd71} with some modifications.  
The following assumptions are used: synchronous rotation of the components, 
the components radiate as black bodies, an orbital period of  2.267460 days 
and a zero epoch of  2448500.76777 \citep{hip}, and the luminosity of the 
secondary component is computed as a function of temperature and surface 
area of that component.  The logarithmic limb darkening law was used. 
Limb darkening coefficients used in the models 
are taken from the tables of \citet{vanhamme93} for stars of the 
appropriate mass and temperature.  The gravity brightening and reflective 
albedo coefficients used are the ones suggested by \citet{wilson01} for 
hot stars with radiative envelopes.  While the temperature of the secondary 
component is allowed to float, the temperature of the primary component 
is held constant at 22000 K, which is consistent with the observed colors 
of the system and the mass of the primary component.

Curve dependent weights were applied in order to compensate for the 
differences in precision between the light and velocity data sets.  The 
\citet{wilson01} code allows the weighting of curves by the standard 
deviation of the curve ($\sigma_c$).  The method of \citet{kallrath99} using
the differences between pairs of consecutive points in phase space was 
used for calculating the value of $\sigma_c$ for each light curve.  In the 
case of the radial velocity curves, the values were not spaced close enough 
to evaluate differences between consecutive points in phase space so the 
truncated FFT method \citep{kallrath99} was used instead by simulating the 
velocity curve to first order with a sine function.  The curve dependent 
weights calculated from this method are listed in Table 5.

The $\sigma_c$ for each velocity curve, when normalized to the curve 
amplitude, is roughly 20 times larger than the $\sigma_c$ of each of the
light curves, which translates to a relative weight in the total solution
on the order of 400 times smaller than the light curve weights.  
Some parameters have no dependence on the light curves (such as the 
semi-major axis and $\Gamma$ velocity).
Full simultaneous solutions of both light and velocity data together
calculated errors for these parameters on order of 50 to 60 percent of 
the parameter values.  
Solutions performed using only the the velocity data produced parameters 
with almost the same values but with errors that were two to four times 
smaller than those obtained in the full simultaneous solution.  This
led to the conclusion that the weights of the velocity curves were probably
understated relative to the light curves.  It seems reasonable that the
velocity data sets should have lower curve weights since they have 
significantly fewer data points.  However, the velocity data is more evenly 
distributed in phase space than any of the light curves.

In order to rectify the difference between the errors in the full solution 
and the solution using only the velocity data, the calculated 
$\sigma_c$ values for the velocity curves were divided by a factor of ten.  
This change gave the velocity curves roughly the same but still less weight 
than each of the light curves.  
When these curve weights were applied to the full simultaneous solution none
of the derived parameter values were significantly different from those 
obtained using the unaltered curve weights.  However, using the altered 
curve weights did reduce the errors of the velocity dependent parameters to 
essentially the same error values given by the solution to the velocity data 
alone.  These corrected $\sigma_c$ values were used in 
the analysis to obtain the best fit parameters.

The \citet{wilson01} code uses a least squares fit that is iterated to 
simultaneously solve the parameters of the system.  The program 
was run in a mode reserved for detached binary systems.
Table 6 lists the best fits to the parameters defining the IT Lib system.  
Figure 2 shows the fit to the radial velocity data and Figure 3 shows the 
fit to the light curve.  The eclipses in the system are partial and the 
system is best modeled by stars of $9.8\pm0.7$ $M_\sun$ and $4.6\pm0.3$ 
$M_\sun$ separated by $17.6\pm1.1$ solar radii.  

\section{Distance from Photometry}

The calculated masses of the components are consistent with main sequence 
stars of spectral class B2 and B7.  The primary component is about three to 
four times as bright as the secondary, so its spectra dominates the observed 
composite spectra.  

The photometric distance to the system can be found knowing the
relative brightness of the components and the absolute magnitude of
the primary component.  It is possible to find the distance
from the absolute magnitude of the combined system but this introduces
additional uncertainties in the form of assumptions that would have to be
made about the absolute properties of both the primary and secondary 
component.  Since the light from the system is dominated by the 
primary, its absolute properties are better constrained.  The relative 
brightness of the components is also well defined by the fit to the 
light curves.  The absolute bolometric magnitude of the primary as 
calculated by the \citet{wilson01} code is -5.20.  The bolometric correction 
for a B2 V star is -2.35 \citep{allen} which yields an absolute V 
magnitude of -2.85 for the A component.  

The apparent V magnitude of the primary ($V_1$) can be obtained from the 
apparent magnitude of the combined system ($V_c$) and the fraction of light 
from the primary relative to the combined system ($F$).

$V_1=V_c+2.5\times log_{10}(1+F)$

The calculation is carried out at maximum light in Johnson V (phase 0.271).
The observed maximum apparent V magnitude of the combined system taken from 
\citet{hip} is 9.09. The apparent magnitude 
comes from the maximum apparent magnitude of the system in the Tycho V 
filter, corrected to Johnson V by the method prescribed by \citet{hip}.
The fraction of light from the primary relative to the combined system (F)
is 0.7767 as calculated by the \citet{wilson01} code.  This yields a
apparent V magnitude of 9.71 for the primary alone.
The extinction factor ($A_V$) from \citet{schelgel98} is 0.702 (assuming
$A_V$/E(B-V)=3.1).  From the above quantities a distance of 
2.4 kpc is obtained which places IT Lib 1.0 kpc above the galactic plane.

There are many sources of error to consider including the uncertainty in the 
physical parameters of the primary component, uncertainty
in the extinction factor, and uncertainty in the relative brightness of
the components.  An error of a few tenths of a magnitude would translate 
into a twenty to thirty percent error in distance.  The true errors in 
this case are unknown so a standard photometric distance error of $30\%$ is 
assumed in the following analysis.

\section{Analysis of Age versus Kinematics}

Knowing the distance, the systemic radial velocity, and the proper motions 
allow one to compute the true space velocity of the system.  Proper 
motions for this system are present in Hipparcos \citep{hip}, the ACT2000 
\citep{act2000}, the Tycho-2 catalog \citep{tycho2}, and the UCAC1 
\citep{ucac1}.  These independent measures were averaged together by the 
method of \citet{martin98} to obtain a more accurate average proper motion 
of $\mu_\alpha=1.82\pm0.67$ mas/yr and $\mu_\delta=-1.26\pm0.76$ mas/yr.  
The systemic radial velocity ($-50.9\pm7.0$ km/s) was taken from the 
analysis of the radial velocity curves.  

The UVW space velocities were computed by the method of \citet{eggen61}.  
The UVW velocities were converted to $V_\pi$, $V_\theta$, $V_z$ assuming 
the solar motion of ($V_\pi$, $V_\theta$, $V_z$) = (-9 km/s, 232 km/s, 7 km/s) 
for an object in close proximity to the sun with (U,V,W)=(0,0,0) 
\citep{mihalasbin81}.  IT Lib{\arcmin}s true space velocity was computed to 
be ($V_\pi$, $V_\theta$, $V_z$) = 
($-1.6\pm4.1$ km/s, $207.0\pm13.3$ km/s, $-35.9\pm7.7$ km/s).  
Errors are difficult to estimate for these velocities because the true 
error in the distance in unknown.  For this analysis the distance error 
is assumed to be $30\%$ ($\pm0.7$ kpc), which is reasonable for a 
photometric distance.  
$V_\pi$ are $V_\theta$ are consistent with a 
star within the rotating galactic disk.  $V_z$ and the height above 
the plane lead to the conclusion that while components of IT Lib{\arcmin}s 
velocity are disk-like that it has been some how ejected out of the 
disk into the galactic halo.  Of particular interest is that IT Lib is 
moving back toward the plane of the galaxy, presumably traveling on the 
down-hill side of a ballistic trajectory.  

The average main-sequence lifetime of a ten solar mass star is about 22 
million years \citep{iben67}.   The orbital integrator of \citet{harding01} 
was used to estimate the number of years that have elapsed since IT Lib 
was last in the galactic plane (Z=0).  By integrating the orbit backwards 
in time it is found that it took $33\pm4$ million years for IT Lib to 
travel from Z = 0 to its present location.  The maximum Z distance that IT 
Lib reaches from the disk on this trajectory is just over 1.1 kpc.  We shall 
return to this apparent discrepancy after a discussion of the abundance
analysis of the primary component.

\section{Chemical Abundance Analysis}

The spectrum used for the abundance analysis of IT Lib was taken at 
phase 0.829 (March 31, 1999, see Table 1).  According to the model of the 
system, at this phase both stars are un-obscured so that the observed 
spectra is a straight blend of the light from both components.  At this 
phase the lines of the primary and secondary component are near maximum 
separation, making it easier to measure the contribution to the line from 
each individual component.  Equivalent widths of twelve lines identified 
with the primary component were measured.  Because both stars are 
un-obscured at this phase the continuum contribution from the secondary 
could easily be removed from the measured equivalent widths.  The continuum 
contribution of the secondary component was removed from the measured 
equivalent widths by calculating the ratio between the continuum flux of 
the primary versus the total combined flux of the system (F) and applying this 
ratio to the equivalent width in this manner:\\
\begin{displaymath}
W{\arcmin} = W\times(1/F),\\
\end{displaymath}
The value of F  is calculated for the appropriate band-pass and phase using 
the \citet{wilson01} code (F=0.816 at phase 0.829).  W is 
equivalent width measured from the composite spectra.  And W{\arcmin} is 
the re-scaled equivalent width used in the abundance analysis.  
  
There were no lines (other than the \ion{H}{1}, \ion{He}{1}, and 
\ion{Mg}{2} 4481 lines) that could be positively identified and 
attributed to the secondary.  This is not surprising due to a combination 
of the signal to noise of the observation, the rotational broadening of 
the components, and the fact that the light from the primary component 
dominates the spectra.

The lines that were measured for the primary component are listed in 
Table 7 along with the atomic data for those lines.   The computer program 
LINES \citep{sneden73,lines} was used to calculate LTE abundances from the 
equivalent widths for individual lines.  The primary broadening mechanism 
in these stars is the quadratic stark effect, for which this version of 
LINES uses the method of \citet{cowley71}.  ATLAS9 \citep{atlas9} was used 
to build the stellar atmospheric models.

Because the primary is slightly distorted the gravity varies by a small 
amount over the surface.  The mean surface gravity for the primary 
component as calculated by the Wilson (2001) code is 
log(g) = 3.75.  The effective temperature of a $9.8 M_\sun$ star is 
approximately 22000 K.  The temperature/gravity relations of \citet{nap93} 
obtain the following values for the Str\"{o}mgren photometry published by 
\citet{hauck98}: T = 22060 K, log(g) = 3.62.  This agrees well with the 
parameters obtained from the model of the system.

The microturbulence was determined using the lines of the $4550 \AA$ 
\ion{Si}{3} triplet.  The relative GF values of the triplet are well known 
and the lines cover a range of equivalent widths making this triplet ideal for 
determining microturbulence.  The microturbulence is calculated from the
triplet by finding the value at which each line in the triplet gives the same
abundance value so that there is no trend in abundance with respect to 
equivalent width.  The microturbulence found for each model used is listed
in Table 8.  These are rather typical values for a B-star of this 
temperature under the assumption of LTE.

The results of the analysis over a range of temperatures and gravities are 
presented in Table 8.  The favored model is 22000 K, 
log(g) = 3.75.  The results for individual lines in the favored model 
are listed in Table 7.  The results using other models over a range of 
temperatures and gravities are presented to demonstrate how differences 
in the atmospheric models effect the abundances.  For comparison the solar 
abundance values \citep{grevesse} and the average abundance values for a 
sample of Population I B-stars \citep{martinphd} are given.  The sample of 
Population I B-Stars are twelve stars situated in the galactic plane 
covering the range of atmospheric parameters 
representative of B-stars.  The abundances in these stars were analyzed 
in the same manner as IT 
Lib.  The abundances of the primary component are 
consistent with those for a normal Population I B-Star over the entire 
range of temperatures and gravities explored.  \ion{N}{2} 
and \ion{Al}{3} are a bit depressed; however, when the errors in the analysis 
are considered these anomalies are negligible.

\section{Discussion}

\subsection{Main Sequence Lifetime versus Travel Time}

The discrepancy between the main-sequence lifetime of the primary 
component (22 Myr) and the travel time back to the disk (33 Myr) may 
present a conundrum.  Three possibilities exist to resolve the apparent 
conflict of facts: the travel time and main-sequence lifetime are not in 
conflict when all possible systematic errors are considered, the A 
component of the IT Lib system is somehow enjoying a prolonged 
main-sequence lifetime due to mass transfer or other method of rejuvenation, 
or the IT Lib system did not form in the galactic disk.

The error budget for calculating the orbit of IT Lib does not allow for 
the possibility that IT Lib was in the galactic plane 22 million 
years ago or less.  Each component of the true space velocity is the sum 
of a term including the radial velocity and a term with a linear 
dependence on the distance to the star.  The radial velocity term is well 
determined compared to the second term which is perturbed by the error in 
the proper motion and multiplied by the uncertainty in the distance.  
Recall that the error in the distance is unknown, so in the Monte-Carlo 
error simulation a distance error of $30\%$ was assumed.  A change in the 
distance would effect the U, V, and W velocities in a linear manner by 
-4.9 km/s/kpc, 1.6 km/s/kpc, and -9.2 km/s/kpc.  If IT Lib is further away 
then it takes even longer for it to journey from the galactic plane to 
its present location so we are only concerned with the possibility that 
the distance has been overestimated.  Because of the way that the 
velocity vector of IT Lib is oriented with respect to the plane of the sky, a 
large change in distance does not have a significant effect on the 
velocity vector of the star.  If the distance to IT Lib is taken to be 
only 1 kpc then it still takes about 28 million years to return to the 
galactic plane.  The likelihood is that there is not enough room in the 
error budget of the true velocity to put IT Lib in the plane any sooner 
than 28 million years ago.

The galactic potential model and integrator used to calculate the 
trajectory of IT Lib are described in detail in \citet{harding01}.  
This model contains disk, halo, bulge, and dark halo components which 
are able to reproduce the observed motions of stars both in the disk 
and the halo under the most rigorous testing.
The model assumes a smooth distribution of matter in the 
galactic disk which is a good approximation for our purposes since the 
gravitational influence of inhomogeneities in the disk only becomes 
significant within a few hundred parsecs of the galactic plane.  Stars on
orbits that take them kiloparsecs out of the plane spend very little of their
time that close so for all intents and purposes the potential they experience
is not significantly perturbed from the smooth approximation.
It is reasonable to assume that this is a robust model that 
does not significantly contribute to the errors associated with 
the travel time back to the disk.

There is probably more uncertainty in the main-sequence lifetime of the 
primary component than in the travel time back to the disk.  The 
main-sequence lifetime of a star is a simple relation between the 
luminosity of the star and the amount of hydrogen available to undergo 
fusion in the core.  Mixing (by rotation, convection, or other means) 
can lengthen the main sequence lifetime of a star by increasing the 
amount of hydrogen available for burning in the core.  Modeling the 
evolution of massive stars becomes a complicated problem requiring the 
accounting of many processes, some of which are only known to an order 
of magnitude \citep{heger00}.  \citet{talon97} demonstrated for a model 
of a 9 $M_\sun$ star the time to hydrogen depletion in the core can vary as 
much as 4 million years just by including or excluding convective 
overshooting from the model.  Stellar rotation also influences mixing.  
\citet{heger00a} calculated that a 10 $M_\sun$ star rotating at 200 km/s has 
a main sequence lifetime of  25 million years versus only 19 million 
years for the same model without rotation.  In the IT Lib system there 
is the additional consideration of the mixing effects that may be induced in 
one component through tidal interaction with the other component.  It is 
difficult to estimate the effect of all these factors on the main-sequence 
lifetime of IT Lib A but it seems reasonable to assume that the 
main-sequence lifetime falls in the range from 22 to 32 million years.

Taking all factors into account it would be reasonable to say that IT Lib 
was formed in the galactic disk just over 30 million years ago and that the
primary component is still on the hydrogen burning main sequence. 
When the uncertainty in the main sequence lifetime of a 10 $M_{\sun}$ 
star in a close binary system is combined with uncertainty in the travel 
time back to the disk this becomes a plausible explanation for IT Lib.

\subsection{Ejection From The Disk And Rejuvenation Through Mass Transfer}

\citet{hoogerwerf01} cite two mechanisms that are thought to produce a 
majority of run-away O and B-type stars: the binary-supernova scenario 
(BSS) and the dynamical ejection scenario (DES).  Stars ejected by the 
DES mechanism should not show any discrepancy between their actual and 
apparent ages.  The problem remains of how one might eject a binary star 
system from the disk.  While ejection scenarios may favor single stars 
it still is possible, however unlikely it may be, that close binaries 
could also be ejected.  IT Lib is a unique system and may well be one of 
those rare binaries that undergo DES ejection.

Stars ejected from the disk by BSS once had a close companion, which 
exploded in a supernova.  When the shell from the explosion passes the 
secondary, the system becomes unbound and the companion flies off with a 
velocity comparable to its original orbital velocity.  Because mass 
transfer would occur in such a system prior to and as a result of the 
supernova explosion the ejected companion is ``rejuvenated.''  
\citet{hoogerwerf01} show by tracing BSS candidates back to their star 
cluster of origin that run-aways produced by BSS are probably 
blue-stragglers, appearing younger than they really are.  If IT Lib were 
ejected by BSS then it would have probably originated in a ternary system 
with the current components orbiting the more massive supernova progenitor 
at a distance greater than the distance between the remaining components.  
The mass transfer from the supernova progenitor to the remaining components 
could possibly explain the discrepancy between their main-sequence 
lifetimes and travel time back to the disk.

There is probably not any appreciable mass transfer occurring in the system 
now because the model shows it is fully detached and there are no spectral 
emission features (particularly $H\alpha$) that could be associated with a 
mass stream or accretion disk.  It is probable that a mass transfer event 
large enough to influence the evolution of either component in the system 
would manifest itself as photospheric abundance anomalies.  In order to move 
enough mass to have a significant impact one would have to dredge deep 
enough to involve processed material in the mass transfer that would end 
up in the outer layers of the component receiving the transfer.

\subsection{In Situ Halo Star Formation}

Another possibility is that the IT Lib system did not form in the galactic 
disk at all.  The apex of the trajectory followed by IT Lib occurred 
10 million years ago at a height of 1.1 kpc.
Assuming that IT Lib formed in a high velocity cloud (HVC) or 
intermediate velocity cloud (IVC) at that time there 
would be no conflict between the main-sequence lifetime of the primary 
and the travel time from its position at birth to its current location.  
The motions of molecular clouds are influenced by magnetic and 
hydrodynamic forces that do not affect the motions of individual stars 
\citep{bm98}.  As a result it seems possible to form a close binary
pair and drop it out of a cloud on a trajectory toward the disk without 
invoking any binary star ejection mechanism.

\citet{vanwoerden93} and \citet{christodoulou97} report that the 
conditions thought to exist in HVCs are conducive to the formation of 
individual massive stars.  Even though the conditions appear to exist in 
the HVCs for star formation to theoretically occur, no observational 
evidence of star formation in these clouds has been found 
\citep{willman02,ivezic97,hambly96}.  Star formation apparently is occurring 
under what are thought to be similar environmental conditions in the HI bridge 
between the LMC and SMC \citep{christodoulou97}.  \citet{comeron01} report 
signs of star formation occurring in the halo of edge-on spiral 
galaxy NGC 253, which they conclude shows it is possible for star formation 
to currently be occurring in the HVCs of the Milky Way.  While the absence 
of evidence is not evidence of absence, it would be easier to accept the 
scenario of IT Lib forming in an HVC or IVC if there was other, more 
compelling, evidence.  Due to uncertainties in the distances and 
trajectories of the IVCs and HVCs it is not possible to determine if 
any known HVC or IVC clouds intersect the orbit of IT Lib at any time 
in the past. 

\subsection{Evidence From Abundances}

The results of the abundance analysis could provide additional information 
that would favor one scenario over another.  If IT Lib were rejuvenated 
and ejected in a BSS ejection then it is expected that the elemental 
abundances observed in the photospheres of the components should be 
enhanced, particularly those abundances of helium and nitrogen 
\citep{hoogerwerf01,heger00a}.  Likewise, if IT Lib had undergone mass 
transfer between components in its past then the abundances might reflect 
this activity.  If IT Lib were formed in an HVC or IVC we would expect 
the abundances of the components to be compatible with those observed by 
\citet{wakker01} for those clouds. 

The analysis of the lines associated with the primary component of IT Lib 
show abundances that are in line with those expected for a 
Population I B-star found in the galactic disk.  Nitrogen does not appear 
to be significantly enhanced as would be expected in the BSS scenario.  
There are no other obvious abundance anomalies as would be expected if 
processed material had been accreted onto the primary.  Even if the 
atmospheric parameters of the primary are varied over a range of gravities 
and temperatures, the derived photospheric abundances are consistent with 
what would be expected for a ``normal'' Population I B-star found in the 
galactic disk.  The abundances of the primary do not lend weight to any of 
the scenarios involving rejuvenation of the system through mass transfer 
nor does IT Lib A appear to be markedly metal poor as might be expected 
from a star formed in an HVC or IVC in the galactic halo.  

\subsection{Conclusion}

There does seem to be enough evidence to suggest that there is 
actually no discrepancy between the main-sequence lifetime of the primary 
and the travel time back to the disk.  When accounting for possible 
errors, particularly those associated with the main-sequence 
lifetime of a massive star, it becomes evident that there is probably no 
discrepancy at all.  It seems most likely that IT Lib is a pair of 
Population I B-stars that were ejected from the star forming regions of 
the galactic disk by DES about 30 million years ago.

It is important to recognize that conclusions reached about the IT Lib system 
do not automatically apply to any other high galactic latitude B-Stars.  
The careful analysis of full space velocities (radial velocities and proper 
motions) using a comprehensive model of the galactic potential will help
identify more Population I B-Stars that have been ejected from the disk.  
The reasons to study a sample of normal Population I B-stars at
large distances may not be obvious, when their nearby cousins are easier 
to study at high precision.  Population I B-Stars that have been ejected from
the disk have a lower limit set on their ages by the analysis of their
trajectory, constraining their ages better than their disk-bound counter
parts.  As a result these stars might be used to start to piece together
an observational picture of the evolution of high mass B-stars on the
main-sequence.  This would be of great benefit to those that model
these stars and may help enhance our understanding of stellar energy 
generation and energy transport processes in general.

\section{Acknowledgments}
I wish to acknowledge R. Earle Luck for his continued assistance 
and support.  I also wish to acknowledge John Feldmeier and Chris Mihos for 
their efforts obtaining photometric observations of IT Lib and John Feldmeier 
for useful conversations concerning this research and reducing the 2.1-m data.
Finally, I want to thank the anonymous referee for the time and effort they
gave to contribute many insightful comments and suggestions that had a 
positive impact on this paper. This work was supported in part by the Jason 
J. Nassau Scholarship Fund and the Townsend Fund.




\clearpage

\figcaption[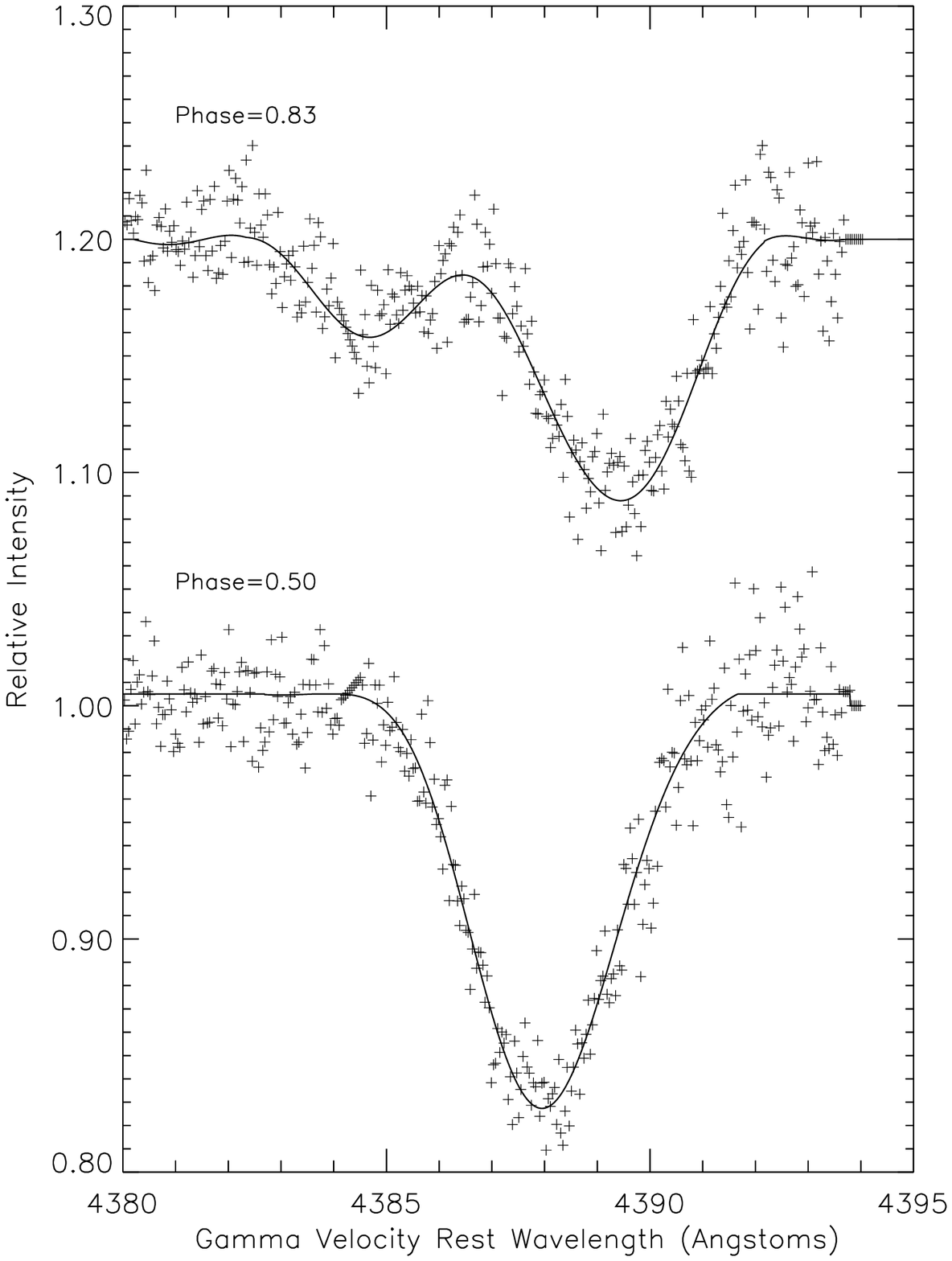]{A spectral plot of the \ion{He}{1} 4487.9 line at phases 0.83 (top) and 0.50 (bottom).  Wavelengths are in the rest frame for the center of mass of IT Lib. The crosses are observed relative flux and the plotted lines are the FFT to the data.\label{fig1}}

\figcaption[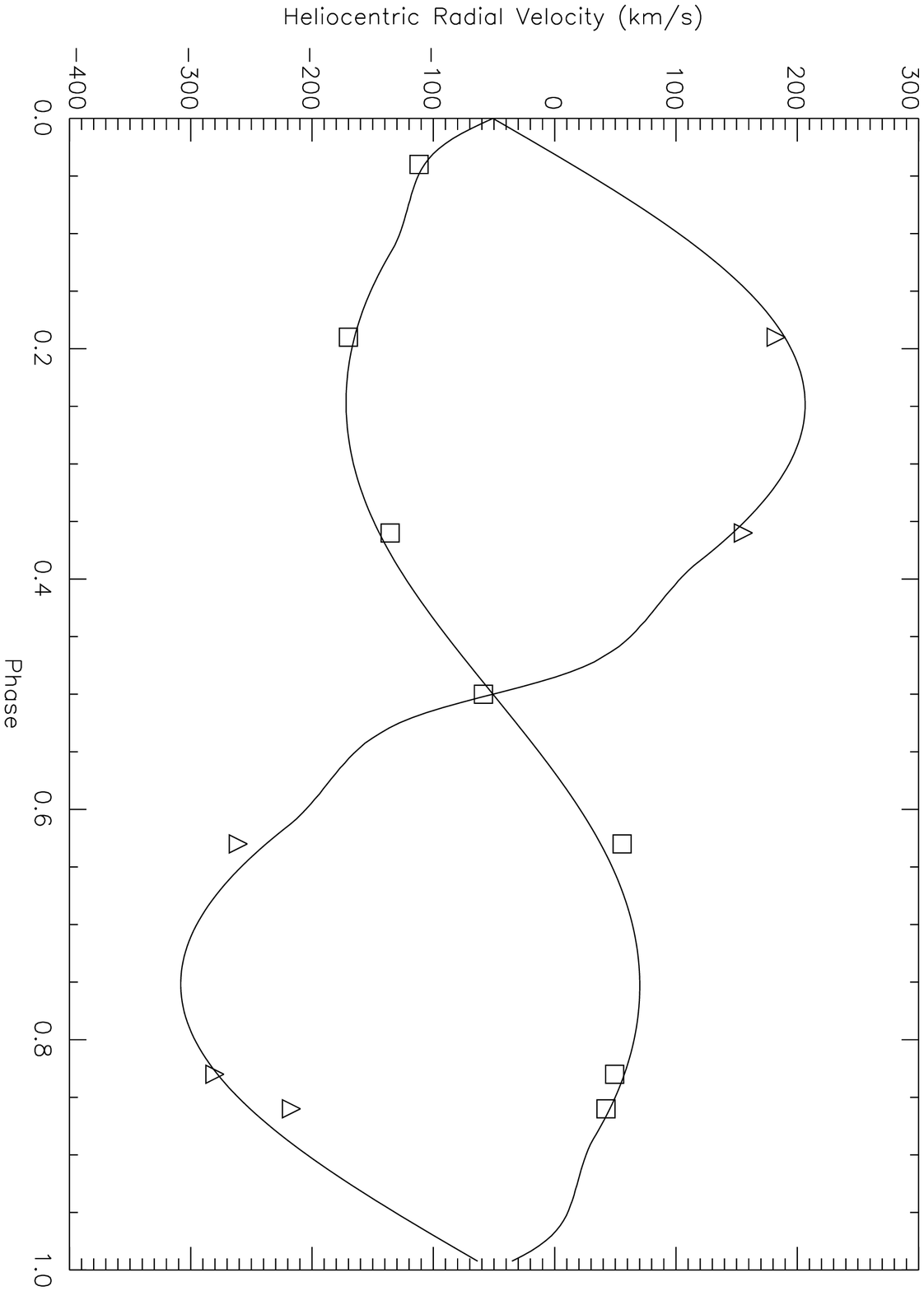]{A plot of the observed radial velocities of the primary (squares) and secondary (triangles) components corrected to the heliocentric rest frame versus phase.  The curves are the predicted line center velocities from the best fit model to the IT Lib system.  The symbols are about the size of the estimated errors.}

\figcaption[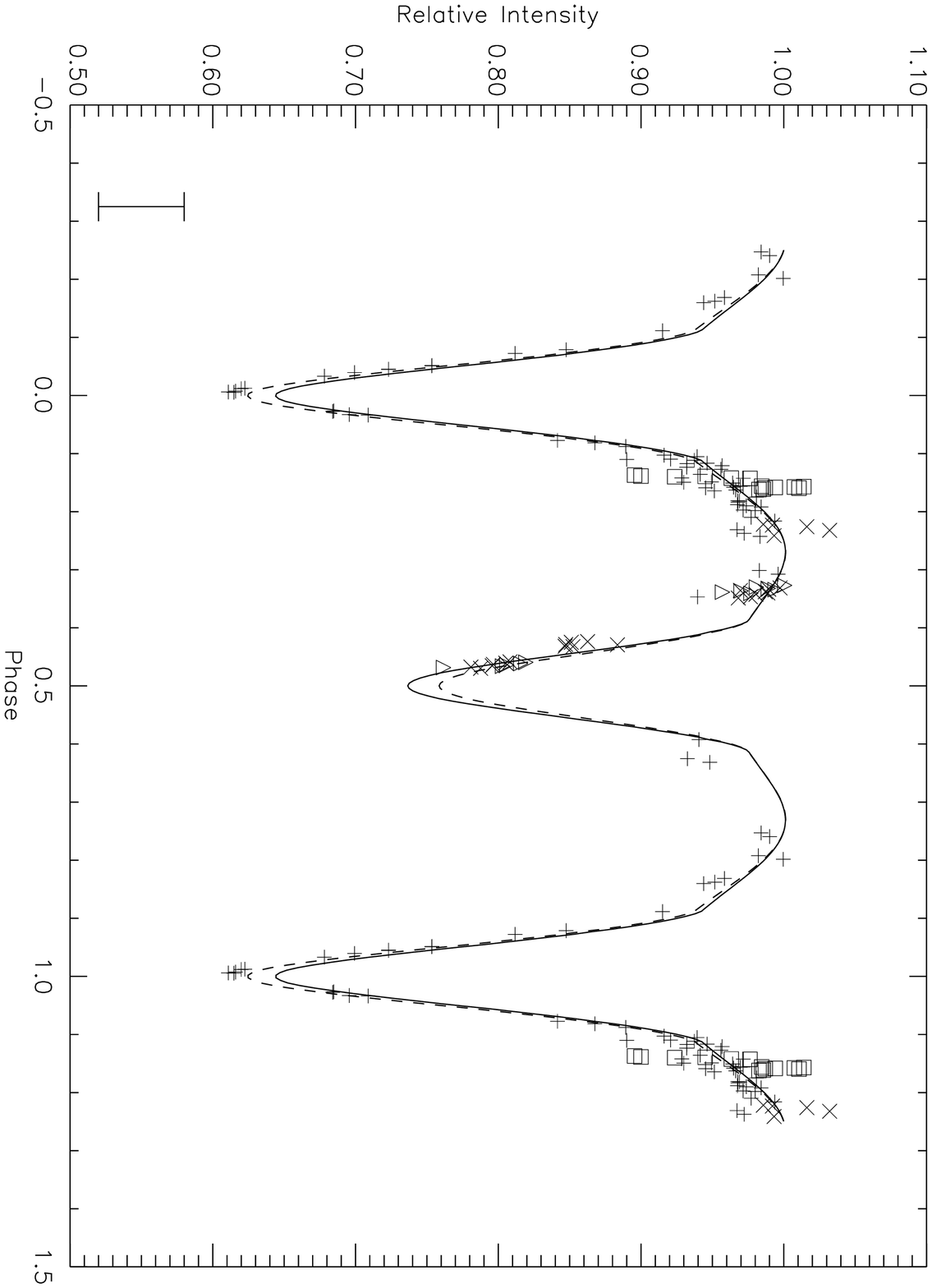]{A plot of the photometric observations used to determine the best fit parameters for the IT Lib system including:  Hipparcos Photometry Annex data \citep{hip} (crosses), data provided by \citet{feld02}(squares), and data obtained on the Nassau 36" reflector in Cousins R (triangles), and Cousins I (x's).  The solid curve is the modeled light curve in Cousins I (deepest in the secondary eclipse).  The dashed curve is the modeled light curve in Johnson V (deepest in the primary eclipse).  At lower left is a error bar representative of the average error $\pm0.03.$}

\clearpage

\begin{deluxetable}{rrrr} 
\tablewidth{0pt}
\tablecolumns{4}
\tablecaption{Summary of Spectroscopic Observations}
\tablehead{
\colhead{Obs Date (UT)}&\colhead{Julian Date}&\colhead{Spectral Coverage}&\colhead{S/N}}
\startdata
March 28, 1999 & 2451265.941 & 4120\AA - 4640\AA & 60 \\
March 31, 1999 & 2451268.949 & 4120\AA - 4640\AA & 65 \\
February 6, 2001 & 2451946.999 & 5590\AA - 6930\AA & 100\\
February 9, 2001 & 2451950.005 & 5590\AA - 6930\AA & 60\\
February 10, 2001 & 2451951.014 & 5590\AA - 6930\AA & 100\\
May 22, 2002 & 2452416.764 & 4850\AA - 5630\AA & 50\\
May 25, 2002 & 2452419.750 & 4850\AA - 5630\AA & 100\\
\enddata
\end{deluxetable}

\begin{deluxetable}{rrrr}
\tablewidth{0pt}
\tablecolumns{4}
\tablecaption{Heliocentric Radial Velocity Measurements}
\tablehead{
\colhead{}&\colhead{}&\multicolumn{2}{c}{Component Velocity}\\
\cline{3-4}\\
\colhead{Julian Date}&\colhead{Phase}&\colhead{A (km/s)}&\colhead{B (km/s)}}
\startdata
2452416.764&0.04&-111.8& \\
2451950.005&0.19&-170.1&182.6\\
2452419.750&0.36&-135.6&155.0\\
2451265.941&0.50&-58.7&\\
2451951.014&0.63&55.6&-260.8\\
2451268.949&0.83&49.5&-280.1\\
2451946.999&0.86&42.3&-217.0\\
\enddata
\end{deluxetable}

\begin{deluxetable}{rrcrr}
\tablecolumns{4}
\tablewidth{0pt}
\tablecaption{New Photometric Observations Used in the Analysis}
\tablehead{
\colhead{Phase}&\colhead{Julian Date}&\colhead{Filter}&\colhead{$\Delta$ Mag\tablenotemark{a}}}
\tablenotetext{a}{$\Delta$ Mag is the magnitude difference with respect to the magnitude of IT Lib in that filter at phase 0.25}
\startdata
0.137&2452348.958&B&$ 0.114\pm 0.027$\\
0.138&2452348.960&B&$ 0.120\pm 0.045$\\
0.139&2452348.963&B&$ 0.114\pm 0.023$\\
0.140&2452348.965&B&$ 0.062\pm 0.024$\\
0.141&2452348.966&B&$ 0.086\pm 0.041$\\
0.143&2452348.971&B&$ 0.041\pm 0.009$\\
0.143&2452348.973&B&$ 0.026\pm 0.024$\\
0.157&2452349.003&B&$ 0.017\pm 0.016$\\
0.158&2452349.005&B&$-0.015\pm 0.034$\\
0.158&2452349.006&B&$-0.008\pm 0.030$\\
0.159&2452349.008&B&$ 0.006\pm 0.032$\\
0.160&2452349.009&B&$ 0.014\pm 0.029$\\
0.161&2452349.012&B&$-0.011\pm 0.049$\\
0.161&2452349.013&B&$ 0.018\pm 0.054$\\
0.162&2452349.015&B&$ 0.016\pm 0.031$\\
0.163&2452349.016&B&$ 0.019\pm 0.036$\\
0.222&2452471.593&I&$ 0.016\pm 0.022$\\
0.223&2452471.596&I&$ 0.009\pm 0.021$\\
0.226&2452471.603&I&$-0.017\pm 0.042$\\
0.232&2452471.617&I&$-0.034\pm 0.050$\\
0.327&2452444.622&R&$-0.001\pm 0.041$\\
0.330&2452444.629&R&$ 0.021\pm 0.043$\\
0.331&2452444.632&I&$ 0.003\pm 0.022$\\
0.332&2452444.634&R&$ 0.012\pm 0.034$\\
0.334&2452444.637&I&$ 0.011\pm 0.022$\\
0.334&2452444.639&R&$ 0.007\pm 0.032$\\
0.336&2452444.642&I&$ 0.034\pm 0.015$\\
0.336&2452444.644&R&$ 0.033\pm 0.028$\\
0.338&2452444.647&I&$ 0.014\pm 0.013$\\
0.339&2452444.648&R&$ 0.048\pm 0.028$\\
0.340&2452444.652&I&$ 0.012\pm 0.023$\\
0.341&2452444.653&R&$ 0.025\pm 0.021$\\
0.342&2452444.657&I&$ 0.021\pm 0.021$\\
0.347&2452444.667&I&$ 0.026\pm 0.047$\\
0.349&2452444.671&I&$ 0.035\pm 0.050$\\
0.459&2452442.653&I&$ 0.231\pm 0.023$\\
0.459&2452442.655&R&$ 0.216\pm 0.040$\\
0.461&2452442.658&I&$ 0.235\pm 0.023$\\
0.462&2452442.660&R&$ 0.221\pm 0.033$\\
0.463&2452442.663&I&$ 0.247\pm 0.015$\\
0.464&2452442.665&R&$ 0.234\pm 0.034$\\
0.465&2452442.668&I&$ 0.248\pm 0.026$\\
0.466&2452442.670&R&$ 0.238\pm 0.028$\\
0.467&2452442.673&I&$ 0.269\pm 0.028$\\
0.468&2452442.675&R&$ 0.296\pm 0.030$\\
0.470&2452442.679&I&$ 0.259\pm 0.029$\\
0.472&2452442.684&I&$ 0.251\pm 0.051$\\
\enddata
\end{deluxetable}

\begin{deluxetable}{lrrr}
\tablecolumns{4}
\tablewidth{0pt}
\tablecaption{Photometric Reference Stars}
\tablehead{
\colhead{}&\colhead{RA Offset\tablenotemark{a}}&\colhead{Dec Offset\tablenotemark{a}}&\colhead{}\\
\colhead{Star}&\colhead{(arc min)}&\colhead{(arc min)}&\colhead{Red Mag\tablenotemark{a}}}
\tablenotetext{a}{Positions and magnitudes from \citet{usnoa2}}
\startdata
0600-18258994&-3.54&2.38&12.4\\
0600-18267326&-1.24&1.63&12.7\\
0600-18266249&0.41&2.75&14.0\\
0600-18259671&-3.17&1.55&14.1\\
0600-18262147&-1.77&0.34&14.1\\
0600-18262375&-1.66&-1.53&13.8\\
0600-18274971&4.92&-3.13&11.8\\
0600-18271754&3.23&0.35&13.3\\
0600-18261506&-2.14&-2.91&14.1\\
0600-18287726&-4.23&-5.00&12.4\\
\cline{1-4}\\
IT Lib&0.00&0.00&9.0\\
\enddata
\end{deluxetable}

\begin{deluxetable}{lc}
\tablecolumns{2}
\tablewidth{0pt}
\tablecaption{Curve Dependent Weights}
\tablehead{
\colhead{Curve}&\colhead{Weight ($\sigma_c$)\tablenotemark{a}}}
\tablenotetext{a}{For the light curves in units of normalize light at phase 0.25}
\startdata
Hipparos V LC&0.0023\\
Johnson B LC&0.0057\\
Cousins R LC&0.0056\\
Cousins I LC&0.0017\\
A component RVC&5.85 km/s\tablenotemark{b}\\
B component RVC&11.93 km/s\tablenotemark{b}\\
\enddata
\tablenotetext{b}{The values used in the analysis were 1/10 these values.  See text for explanation.}
\end{deluxetable}

\begin{deluxetable}{lcc}
\tablecolumns{3}
\tablewidth{0pt}
\tablecaption{Parameters Derived for IT Lib System}
\tablehead{
\colhead{Parameter}&\multicolumn{2}{c}{Value}}
\startdata
semi-major axis&\multicolumn{2}{c}{$17.6\pm1.1 R_\sun$}\\
$\Gamma$ velocity&\multicolumn{2}{c}{$-50.9\pm7.0 km/s$}\\
inclination&\multicolumn{2}{c}{$78.1\pm3.5 deg$}\\
$\Omega_1$&\multicolumn{2}{c}{$3.11\pm0.24$}\\
$\Omega_2$&\multicolumn{2}{c}{$3.03\pm0.16$}\\
$T_2$&\multicolumn{2}{c}{$13620\pm2068 K$}\\
$M_2/M_1$&\multicolumn{2}{c}{$0.467\pm0.044$}\\
eccentricity&\multicolumn{2}{c}{$-0.0095\pm0.0129$}\\
\cline{1-3}\\
\colhead{}&\multicolumn{2}{c}{Component}\\
\colhead{Parameter}&\colhead{A}&\colhead{B}\\
\cline{1-3}\\
Mass&$9.77\pm0.65 M_\sun$&$4.57\pm0.30 M_\sun$\\
Mean Radius&$6.87R_\sun$&$4.77R_\sun$\\
$M_{bol}$&-5.20&-2.33\\
Mean log(g)&3.75 cgs&3.74 cgs\\
\enddata
\end{deluxetable}

\begin{deluxetable}{lrrrrrr}
\tablecolumns{7}
\tablewidth{0pt}
\tablecaption{Atomic Line Data}
\tablehead{
\colhead{Atomic}&\colhead{}&\colhead{}&\colhead{}&\colhead{$EW (m\AA)$}&\colhead{$EW (m\AA)$}\\
\colhead{Species}&\colhead{$\lambda_0 (\AA)$}&\colhead{$E_0 (eV)$}&\colhead{gf}&\colhead{measured}&\colhead{scaled}&\colhead{Abund}}
\startdata
N II&4227.750&21.5995&$8.689E-01$&29.1&35.6&7.89\\
N II&4241.800&23.2463&$5.400E+00$&26.9&32.9&7.46\\
\cline{1-7}\\
O II&4351.260&25.6614&$1.687E+01$&40.8&50.0&8.27\\
O II&4416.975&23.4192&$8.375E-01$&93.1&114.0&8.79\\
O II&4596.177&25.6635&$1.585E+00$&73.8&90.4&8.89\\
\cline{1-7}\\
Al III&4512.565&17.8083&$2.624E+00$&19.2&23.5&5.64\\
Al III&4529.189&17.8182&$4.688E+00$&68.3&83.7&6.16\\
\cline{1-7}\\
Si III&4552.622&19.0163&$1.959E+00$&131.7&161.3&7.08\\
Si III&4567.840&19.0163&$1.175E+00$&118.3&144.9&7.15\\
Si III&4574.757&19.0163&$3.926E-01$&69.6&85.2&7.08\\
\cline{1-7}\\
Fe III&4222.271&20.8704&$1.800E+00$&24.3&29.8&7.72\\
Fe III&4310.355&20.8688&$1.519E+01$&21.8&26.7&7.25\\
\enddata
\end{deluxetable}

\begin{deluxetable}{rrrlllll}
\tablecolumns{8}
\tablewidth{0pt}
\tablehead{
\multicolumn{3}{l}{Model}&\multicolumn{5}{c}{Mean Abundances}\\
\cline{4-8}\\
\colhead{$T_{eff}$}&\colhead{log(g)}&\colhead{$v_{turb}$}&\colhead{\ion{N}{2}}&\colhead{\ion{O}{2}}&\colhead{\ion{Al}{3}}&\colhead{\ion{Si}{3}}&\colhead{\ion{Fe}{3}}}
\startdata
22000&4.00&10.0&$7.74\pm0.30$&$8.79\pm0.35$&$5.96\pm0.37$&$7.23\pm0.04$&$7.58\pm0.34$\\
22000&3.75&10.0&$7.67\pm0.30$&$8.65\pm0.34$&$5.90\pm0.36$&$7.10\pm0.04$&$7.49\pm0.34$\\
22000&3.50&10.0&$7.63\pm0.30$&$8.51\pm0.32$&$5.86\pm0.36$&$6.99\pm0.04$&$7.40\pm0.34$\\
22000&3.00&9.0&$7.65\pm0.30$&$8.32\pm0.30$&$5.93\pm0.36$&$6.89\pm0.04$&$7.33\pm0.34$\\
26000&4.00&9.0&$7.71\pm0.33$&$8.30\pm0.29$&$6.11\pm0.36$&$7.02\pm0.04$&$7.43\pm0.37$\\
24000&4.00&9.0&$7.66\pm0.31$&$8.48\pm0.32$&$5.96\pm0.37$&$7.06\pm0.04$&$7.44\pm0.35$\\
20000&4.00&10.0&$7.98\pm0.29$&$9.29\pm0.37$&$6.16\pm0.42$&$7.60\pm0.05$&$7.87\pm0.32$\\
\cline{1-8}\\
\multicolumn{3}{l}{Avg Norm B-Star\tablenotemark{a}}&$7.99\pm0.33$&$8.57\pm0.17$&$6.32\pm0.33$&$7.36\pm0.18$&$7.37\pm0.22$\\
\cline{1-8}\\
\multicolumn{3}{l}{Solar Photosphere\tablenotemark{b}}&$7.92\pm0.06$&$8.83\pm0.06$&$6.47\pm0.07$&$7.55\pm0.05$&$7.50\pm0.05$\\
\enddata
\tablenotetext{a}{\citet{martinphd}}
\tablenotetext{b}{\citet{grevesse}}
\end{deluxetable}



\clearpage

\begin{thebibliography}{}
\bibitem[Bidelman(1948)]{bidelman51} Bidelman, W.P.  1948, \pasp, 60, 264
\bibitem[Binney \& Merifield(1998)]{bm98} Binney, J. \& Merifield, M.  1998, Galactic Astronomy, (Princeton, NJ: Princeton Univ. Press)
\bibitem[Comer\'{o}n et al.(2001)]{comeron01} Comer\'{o}n, F., Torra, J., M\'{e}ndez, R.A., \& G\'{o}mez, A.E.  \aap, 2001, 366, 796
\bibitem[Cowley(1971)]{cowley71} Cowley, C. R.  1971, Obs, 91, 139
\bibitem[Cox(2000)]{allen} Cox, Aurthur N.  2000, Allen's Astrophysical Quantities, Fourth Edition, (New York: Springer)
\bibitem[Christodoulou et al.(1997)]{christodoulou97} Christodoulou, D.M., Tohline, J.E., and Keenan, F.P.  \apj, 1997, 486, 810
\bibitem[Eggen(1961)]{eggen61} Eggen, O.  1961, ROB, 41, E268
\bibitem[Feldmeier \& Mihos(2002)]{feld02} Feldmeier, J. \& Mihos, J.C. 2002, personal communication
\bibitem[Grevesse \& Sauval(1998)]{grevesse} Grevesse, N. \& Sauval, A.J.  1998, Space Science Reviews, 85, 161
\bibitem[Hambly et al.(1996)]{hambly96} Hambly, N.C., Wood, K.,D., Keenan, F.P., Kilkenny, D., Dufton, P.L., Miller, L., Gilmore, G., Irwin, M.J., \& Totten, E.J.  1996, \aap, 306, 119. (also presented at ASP conf \# 92)
\bibitem[Harding et al.(2001)]{harding01} Harding, P., Morrison, H.L., Olszewski, E.W., Arabadjis, J., Mateo, M., Dohm-Palmer, R.C., Freeman, K., \& Norris, J.E. 2001, \aj, 122, 1397
\bibitem[Hauck \& Mermilliod(1998)]{hauck98} Hauck, B \& Mermilliod, M.  1998, \aaps, 129, 431
\bibitem[Heger et al.(2000)]{heger00} Heger, A., Langer, N., \& Woosley, S.E.  2000, \apj, 528, 368
\bibitem[Heger \& Langer(2000)]{heger00a} Heger, A. \& Langer, N.  2000, \apj, 544, 1016
\bibitem[Hill(1970)]{hill70} Hill, P.W.  1970, \mnras, 150, 23
\bibitem[Hoffmeister(1935)]{hoff35} Hoffmeister, C. 1935, Astron. Nach., 255, 405
\bibitem[Hog et al.(2000)]{tycho2} Hog E., Fabricius C., Makarov V.V., Urban S., Corbin T., Wycoff G., Bastian U., Schwekendiek P., \& Wicenec A.  2000, \aaps. 355, 27 (Tycho-2)
\bibitem[Hoogerwerf et al.(2001)]{hoogerwerf01} Hoogerwerf, R., de Bruijue, J.H.J., \& de Zeeuw, P.T.  2001, \aap, 365, 77
\bibitem[Houk \& Smith-Moore(1988)]{mss88} Houk, Nancy \& Smith-Moore, M.  1988, Michigan Catalogue of Two-Dimensional Spectral Types For the HD Stars, Vol 4, (Ann Arbor: University of Michigan)
\bibitem[Humason \& Zwicky(1947)]{humason47}Humason, M.L. \& Zwicky, F. 1947, \apj, 105, 85
\bibitem[Iben(1967)]{iben67} Iben, I.  1967, \araa, 5, 624.
\bibitem[Ivenzic \& Christodoulou(1997)]{ivezic97} Ivezic, Z. \& Christodoulou, D.M.  1997, \apj, 486, 818
\bibitem[Kallrath \& Milone(1999)]{kallrath99} Kallrath, J. \& Milone, E.F.  1999, Modeling and Analysis of Eclipsing Binary Observations, (New Your: Springer-Velag)  
\bibitem[Kazarovets et al.(1999)]{ibvs99} Kazarovets, E.V., Samus, N.N., Durlevich, O.V., Frolov, M.S., Antipin, S.V., Kireeva, N.N., \& Pastukhova, E.N.  1999, IBVS, 4659
\bibitem[Kurucz(1993)]{atlas9} Kurucz, R.  1993, ATLAS9 Stellar Atmosphere Programs and 2 km/s grid. Kurucz CD-ROM No. 13, (Cambridge, Mass: Smithsonian Astrophysical Observatory)
\bibitem[Luck(2001a)]{asp}Luck, R.E. 2001a, ASP spectra analysis software, private communication
\bibitem[Luck(2001b)]{lines}Luck, R.E. 2001b, LINES abundance analysis software, private communication
\bibitem[McCarthy et al.(1993)]{sandiford} McCarthy, J.K., Sandiford, B.A., Boyd, D., \& Booth, J.  1993, \pasp, 105, 881
\bibitem[Martin(2003)]{martinphd} Martin, J.C. 2003, Ph.D Thesis, Case Western Reserve University
\bibitem[Martin \& Morrison(1998)]{martin98} Martin, J.C. \& Morrison, H.L. 1998, \aj, 116, 1724
\bibitem[Mihalas \& Binney(1981)]{mihalasbin81} Mihalas, D. \& Binney, J. 1981, Galactic Astronomy, Second Edition, (San Francisco: Freedman)
\bibitem[Mochejska et al.(2001)]{mochejska01} Mochejska, B.J., Kaluzny, J., Stanek, K.Z., Sasselov, D.D., \& Szentgyorgyi, A.H.  2001, \aj, 121, 2032
\bibitem[Monet et al.(1998)]{usnoa2} Monet, D., Bird, A., Canzian, B., Dahn, C., Gutter, H., Harris, H., Henden, A., Levine, S., Luginbuhl, C., Monet, A.K.B., Rhodes, A., Ripe, B., Sell, S., Stone, R., Vrba, F., \& Walker, R.  1998, The USNO A-2.0 Catalog, (Flagstaff, AZ: US Naval Observatory)
\bibitem[Napiwotzki et al.(1993)]{nap93} Napiwotzki, R., Schonberner, \& Wenske, 1993, \aap, 268, 653
\bibitem[Schlegel et al.(1998)]{schelgel98} Schlegel, D., Finkbeiner, D., \& Davis, M.,  \apj, 1998, 500, 525
\bibitem[Sneden (1973)]{sneden73} Sneden, C.  1973, Ph.D. Thesis, University of Texas
\bibitem[Strohmeier(1967)]{stroh67} Strohmeier, W.  1967, IBVS, 178.
\bibitem[Talon et al.(1997)]{talon97} Talon, S., Zahn, J., Maeder, A., \& Meynet, G.  1997, \aap, 322, 209
\bibitem[Turon et al.(1997)]{hip} Turon, C., Creze, M., Egret, D., Gomez, A., Grenon, M., Jahrei\ss, H., Requieme, Y., Argue, A. N., Bec-Borsenberger, A., Dommanget, J., Mennessier, M. O., Arenou, F., Chareton, M., Crifo, F., Mermilliod, J. C., Morin, D., Nicolet, B., Nys, O., Prevot, L., Rousseau, M., Perryman, M. A. C., et al.  1997, Bull. Inform. CDS, 41, 9. (HIP)
\bibitem[Urban et al.(1997)]{act2000}Urban S.E., Corbin T.E., \& Wycoff G.L.  1997, BAAS, 191, \#57.07. (ACT200)
\bibitem[Van Hamme(1993)]{vanhamme93} Van Hamme, W. 1993, \aj, 106, 2096.
\bibitem[Van Woerden(1993)]{vanwoerden93} Van Woerden, Hugo.  1993, in Luminous High Latitude Stars, ed. D.D. Sasselov, (San Francisco:  ASP), 11
\bibitem[Willman et al.(2002)]{willman02} Willman, B., Dalcanton, J., Ivezic, Z., Schneider, D.P., \& York, D.G.  2002, \aj, in press
\bibitem[Wilson \& Devinney (1971)]{wd71} Wilson, R.E \& Devinney, E. J.  1971, \apj, 166, 605
\bibitem[Wilson(2001)]{wilson01} Wilson, R.E.  2001, Computing Binary Star Observables (Gainesville: Depart. Astronomy, Univ. Florida)
\bibitem[Wakker(2001)]{wakker01} Wakker, B. P.  2001, \apjs, 136, 463.
\bibitem[Wood(1958)]{wood58} Wood, F.B.  1958, \aj, 63, 504
\bibitem[Zacharias et al.(2000)]{ucac1} Zacharias, N., Urban, S.E., Zacharias, M.I., Hall, D.M., Wycoff, G.L., Rafferty, T.J., Germain, M.E., Holdenried, E.R., Pohlman, J.W., Gauss, F.S., Monet, D.G., \& Winter, L..  2000, AJ, 120, 2131
\end{thebibliography}
\end{document}